\documentclass[twocolumn,twoside,superscriptaddress,nofootinbib]{revtex4}

\usepackage[english]{babel}
\usepackage{ngerman}
\usepackage[utf8x]{inputenc}
\usepackage{amssymb}
\usepackage{amsthm,bbm}
\usepackage{amsfonts}
\usepackage[dvips]{color}
\usepackage{epsfig}
\usepackage{dsfont}
\usepackage[usenames,dvipsnames]{pstricks}
\usepackage{amsmath}
\usepackage{graphicx}
\usepackage{mathrsfs}

\graphicspath{{graphics/}}

\definecolor{frametitle.bg}{rgb}{0.6549019607843137,0.6431372549019608,0.996078431372549}
\definecolor{bars.bg}{rgb}{0.17647058823529413,0.14901960784313725,0.996078431372549}

\newcommand{\dg}[1]{\mathrm{dg}(#1)}
\newcommand{\ZZ}{\mathbb{Z}}
\newcommand{\stabgen}{\mathbb{S}}
\newcommand{\NN}{\mathbb{N}}
\newcommand{\Id}{\mathds{1}}
\newcommand{\CN}{\mathcal{N}}
\newcommand{\SCB}{\mathscr{B}}
\newcommand{\SCA}{\mathscr{A}}
\newcommand{\SCC}{\mathscr{C}}
\newcommand{\FA}{\mathfrak{A}}
\newcommand{\tr}{\mathrm{tr}\,}
\newcommand{\sca}{{\bf t}}
\newcommand{\scb}{{\bf b}}
\newcommand{\CQCA}{\mathrm{T}}
\newcommand{\ket}[1]{\left|#1\right>}
\newcommand{\stabilizer}{\mathcal{S}}
\def\fin{\hfill $\lozenge$}
\newcommand{\weyl}[1]{{\bf w}(#1)}
\newcommand{\ul}[1]{\underline{#1}}
\def\varia{{u}}
\newcommand{\scamat}{{\left(\begin{array}{cc}\sca_{11}&\sca_{12}\\\sca_{21}&\sca_{22}\end{array}\right)}}

\newtheorem{thm}{Theorem}[section]
\newtheorem{definition}[thm]{Definition}
\newtheorem{lemma}[thm]{Lemma}
\newtheorem{example}[thm]{Example}

\begin{document}
\begin{abstract}
Clifford quantum cellular automata (CQCAs) are a special kind of quantum cellular automata (QCAs) that incorporate Clifford group operations for the time evolution. Despite being classically simulable, they can be used as basic building blocks for universal quantum computation. This is due to the connection to translation-""invariant stabilizer states and their entanglement properties. We will give a self-contained introduction to CQCAs and investigate the generation of entanglement under CQCA action. Futhermore, we will discuss finite configurations and applications of CQCAs.
\end{abstract}
\title{Entanglement Generation of Clifford Quantum Cellular Automata}
\author{Johannes G\"utschow}
\affiliation{Institut f\"ur Theoretische Physik, Universit\"at Hannover, Appelstra{\ss}e 2, 30167 Hannover \email{johannes.guetschow(at)itp.uni-hannover.de}}
\date{\today}
\maketitle


\maketitle
\section{Introduction}
  In this work we will give an introduction to Clifford quantum cellular automata (CQCAs) \cite{SchlingemannCQCA,Guetschow2009} and study their entanglement generation properties. Quantum cellular automata (QCAs) \cite{Werner2004} are a quantum computational model that only requires global control for the time evolution. The only part of a QCA computation that needs local control is the preparation of the input state, which in the case of a QCA capable of universal quantum computation \cite{Shepherd2006,Raussendorf2005} encodes the program and the data input. But not all QCAs are universal quantum computers. The CQCAs we deal with here are even classically simulable. Yet they are of great use in several quantum computational schemes. Furthermore, the classical structure allows an in-depth analysis of their time evolution, which can give us a good intuition about the properties of general QCAs. Most of the results presented here are also included in \cite{Guetschow2009}. In this work we focus on entanglement generation and try to present the theory of CQCAs in a self-contained manner. For a more detailed and mathematically concise formulation we refer to \cite{SchlingemannCQCA} and \cite{Guetschow2009}.
  
  We will begin with a short introduction to QCAs and CQCAs as well as the classical description of CQCAs. Then we will introduce translation-""invariant pure stabilizer states and study their entanglement in a bipartite setting. Putting the results of the first two sections together, we will investigate the evolution of stabilizer states under CQCA action and derive a simple expression for the generation of entanglement by CQCAs acting on stabilizer states. Namely, the entanglement generation is linear and depends only on the degree of the trace of the polynomial matrix representing the CQCA. We will then make some short remarks on CQCAs over finite chains and finally give a short introduction to applications of CQCAs in quantum computing \cite{Raussendorf2001,Raussendorf2005a,fitzsimons2006}. 

\section{Clifford Quantum Cellular Automata} 
  \subsection{Reversible Quantum Cellular Automata}
\begin{figure*}[tp]
      \begin{center}
        \scalebox{1}{
          \includegraphics{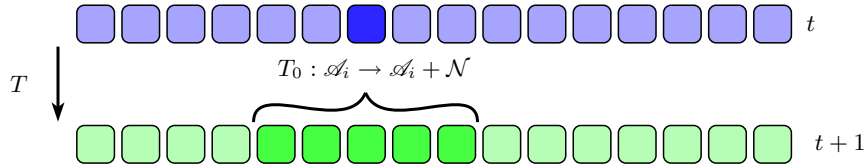}
        }
      \caption{Time evolution of a CQCA. The images of the observables on system $\SCA_i$ are contained in the set of observables of the systems $\SCA_i+\CN$ in the next time step.}
      \label{fig:qca}
      \end{center}
    \end{figure*}

    A reversible quantum cellular automaton (QCA) $\CQCA$ is a translation-""invariant reversible discrete time operation on a translation-invariant lattice of quantum systems. So, in each discrete step of the time evolution of our lattice of quantum systems, we will do the same thing and we will do it in a way independent of the position in our lattice. Because we deal with infinite lattices, e.g. spin chains, it is convenient to look at the time evolution of observables rather than states to circumvent mathematical problems with Hilbert spaces on infinitely many tensor factors. Using the observable evolution, we get the nice property that a QCA is uniquely determined by the image of the observables on a single site \cite{Werner2004}. We call this local operation $\CQCA_i$ if it acts on the observables of the $i$th system. As we defined our QCA to be translation invariant, all $\CQCA_i$ are the same and we pick one, namely $\CQCA_0$, as a representative. We now introduce another property of QCAs: locality. In each time step we only allow a finite propagation of information on the lattice, i.e.\ two states that only differ on one site may only differ on a neighborhood of finitely many sites around the original site after the application of the QCA. This means that the image of the observables at site $i$ is a subset of the the observables of a finite number of sites surrounding site $i$. We call this set the neighborhood $\CN$ of the automaton. This is depicted in Figure~\ref{fig:qca}. 
    
    In mathematical terms we say that a QCA is a local, translation-""invariant automorphism of the algebra of observables $\FA$. An automorphism $\CQCA$ has the property $\CQCA(AB)=\CQCA(A)\CQCA(B),\,\forall A,B\in\FA$. The local operation $\CQCA_0$ is a homomorphism from $\FA_i$ to $\FA(\CN+i)$, where $\CN+i$ denotes the neighborhood of the $i$th cell and $\FA_i$ the observable algebra of the cell. A homomorphism preserves commutation relations in the sense that $[\CQCA(A),\CQCA(B)]=\CQCA[A,B]$ holds. But in our local rule $\CQCA_0$, information about the evolution of neighboring cells does not play a role. However, for $\CQCA$ to be an automorphism, the images of two commuting observables still have to commute after the evolution. So, we impose this condition on $\CQCA_0$ by demanding that all observables in $\CQCA(\FA_i)$ commute with all observables in $\CQCA(\FA_{i+x})$, $x\ne 0$, where $\CQCA(\FA_i)$ denotes the algebra of the images of observables localized on cell $i$.
  \subsection{Clifford Quantum Cellular Automata}
    Now we restrict ourselves to a special kind of QCAs, the Clifford quantum cellular automata (CQCAs), which employ the Clifford group operations. Clifford operations map tensor products of Pauli matrices to tensor products of Pauli matrices times a phase. Thus, the local rule $T_0$ of a CQCA maps single Pauli matrices to tensor products of Pauli matrices times a phase. Furthermore, we only consider one-dimensional lattices, i.e.\ spin chains. We will illustrate this with an example. 
    \begin{example}
      The so-called glider CQCA $\CQCA_G$ has the following local rule:
    \begin{equation*}
      \begin{array}{rcc}
        \CQCA_{G,0} [X_i] &=& Z_i,\\
        \CQCA_{G,0} [Z_i] &=& Z_{i-1}\otimes X_i\otimes Z_{i+1},
      \end{array}
    \end{equation*}
    where $X_i$ and $Z_i$ denote the Pauli matrices at system $i$.
    The image of $Y_i$ follows from the product of the images of $X_i$ and $Z_i$, because we require $\CQCA_0$ to be a homomorphism:
    \begin{equation*}
      \CQCA_{G,0} [Y_i] = -Z_{i-1}\otimes Y_i\otimes Z_{i+1}.
    \end{equation*}
    The global transformation $\CQCA_G$ has to be an automorphism. We already constructed the image of $Y_i$ accordingly. Now we have to check if the commutation and anti-commutation relations between neighboring cells are preserved. We have
    \begin{equation*}
      [\CQCA_{G,0} X^i,\CQCA_{G,0} X^j]=[Z^i,Z^j]=0,
    \end{equation*}
    \begin{eqnarray*}
      &&[\CQCA_{G,0} Z^i,\CQCA_{G,0} Z^j]\\
      &=&[Z^{i-1}\otimes X^i\otimes Z^{i+1},Z^{j-1}\otimes X^j\otimes Z^{j+1}]\\
      &=&0,        
    \end{eqnarray*}
    and
    \begin{eqnarray*}
      [\CQCA_{G,0} Z^i,\CQCA_{G,0} X^j]&=&[Z^{i-1}\otimes X^i\otimes Z^{i+1},Z^j]=0,i\ne j,\\
      \{\CQCA_{G,0} Z^i,\CQCA_{G,0} X^i\}&=&\{Z^{i-1}\otimes X^i\otimes Z^{i+1},Z^i\}=0,
    \end{eqnarray*}
    so our local rule extends to a global automorphism. As the global rule acts in exactly the same way on single-site Pauli matrices as the global rule, we will from now on always use the global rule $\CQCA_G$.
    
    By neglecting a global phase, we can also think of the CQCA $\CQCA_G$ as a classical automaton acting on the labels $(1\,\widehat{=}\,X,2\,\widehat{=}\,Y,3\,\widehat{=}\,Z,0\,\widehat{=}\,\Id)$ of the Pauli matrices. We define the operation $\odot$, which has the following properties to resemble the multiplication of Pauli matrices: $i\odot i=0$, $i\odot j=k$ for $i,j,k=1,2,3$ with $i\ne j\ne k$ and $0\odot i=i\odot 0=i$ for $i=0,\ldots,3$. Now we can illustrate the evolution of the observable $Z^{-1}\otimes Y^0\otimes X^1$ as follows (the underlined labels are situated at the origin.):  
    \begin{eqnarray*}
      \CQCA_G(3\,\ul{2}\,1)&=&\CQCA_G(3\,\ul{0}\,0)\odot \CQCA_G(0\,\ul{2}\,0)\odot \CQCA_G(0\,\ul{0}\,1)\\
      &=&
      \begin{array}{ccccc}
        &3&1&\ul{3}&\\
        \odot&&3&\ul{2}&3\\
        \odot&&&&3\\
        \hline
        =&3&2&\ul{1}&0
      \end{array}
      =(3\,2\,\ul{1})
    \end{eqnarray*}    
    We can see that this observable only moves on the lattice by one step. We call observables with this property ``gliders'' and the automata which allow for such observables glider automata. Figure \ref{fig:glider} gives an explanation for this behavior.
    \begin{figure*}[tp]
      \begin{center}
        \scalebox{1}{
          \includegraphics{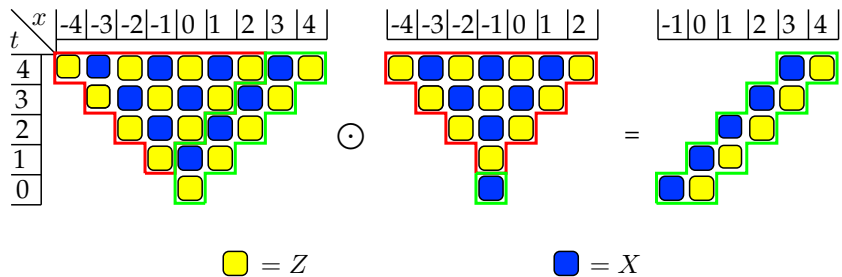}
        }
      \caption{Time evolution of the glider CQCA $\CQCA_G$: On the left side the individual evolutions of single $Z$ and $X$ observables are shown. They both generate ``checkerboards'' which are shifted against each other by one time-step. By shifting the original $X$ also in space we overlap the checkerboards such that they cancel out leaving only a small strip in the space-time picture. This strip is generated by the original observable moving on the spin chain. Such observables are called gliders.}
      \label{fig:glider}
      \end{center}
    \end{figure*}
    \fin
    \end{example}
    \subsubsection{Classical Description of CQCA}
      It is well known that Clifford operations are classically simulable in an efficient way \cite{Gottesman1997}. We use this to derive a classical description of CQCAs, which greatly simplifies our further analysis. Heuristically speaking we replace the CQCAs acting on observables by a classical cellular automaton (CA) acting on the labels of Pauli matrices as we did with the example CQCA in the last section. Mathematically we do the following: first we introduce a function ${\bf w}$ that maps vectors over $\ZZ_2$ to Pauli matrices\footnote{$\ZZ_2$ is the finite field consisting of $0$ and $1$ with addition modulo $2$.}:
      \begin{equation}
        X=\weyl{1,0},\,Y=\mathrm{i}\weyl{1,1},\,Z=\weyl{0,1},\,\Id=\weyl{0,0}.
      \end{equation}
      This way we gave the Pauli matrices classical labels from a vector space. A quantum cellular automaton mapping Pauli matrices to Pauli matrices can now be described as a classical automaton acting on the labels. But, before we introduce the classical description of CQCAs, we have to introduce tensor products of Pauli matrices to our classical description. To do that we extend the function ${\bf w}$ in a way that it maps vectors of binary strings to tensor products of Pauli matrices:
      \begin{equation}
        \weyl{\xi}=\bigotimes_{x\in\ZZ}\weyl{\xi(x)},\,\xi(x)=(\xi_X(x),\xi_Z(x))\in\ZZ_2^2.
      \end{equation}
      The ones in $\xi_X$ account for the Pauli $X$ matrices, the ones in $\xi_Z$ for the $Y$ matrices, and if $\xi_X(x_0)=\xi_Y(x_0)=1$, there is a $Y$ at $x_0$:
      \begin{displaymath}
        {\bf w}\left(\begin{array}{cccccccc}
                      \cdots&0&1&1&0&1&0&\cdots\\
                      \cdots&0&0&1&1&0&0&\cdots
                    \end{array}\right)=\cdots\Id\otimes X\otimes iY\otimes Z\otimes X\otimes\Id\cdots.
      \end{displaymath}
      The binary strings are infinitely long, but, as we always deal with localized observables, only finitely many positions have non-zero entries. We omit the zeros before and after the observable to get a more convenient notation. Similarly, we will omit the identities before and after the observable. The binary strings form a vector space over $\ZZ_2$, which we call phase space.
      
      The addition in the classical description is commutative, so we have to encode the commutation relations in an auxiliary function. Our operators fulfill the condition
      \begin{equation*}
        \weyl{\xi + \eta}=(-1)^{\eta_X\xi_Z}\weyl{\xi}\weyl{\eta},
      \end{equation*}
      which is known from the Pauli matrices as 
      \begin{equation*}
        \sigma_k=\mathrm{i}\varepsilon_{ijk}\sigma_i\sigma_j,
      \end{equation*}
      where $\sigma_1=X$, $\sigma_2=Y$, $\sigma_3=Z$, and $\varepsilon_{ijk}=1$ if $(i,j,k)$ is an even permutation of $(1,2,3)$, $\varepsilon_{ijk}=-1$ if $(i,j,k)$ is an odd permutation of $(1,2,3)$, and $\varepsilon_{ijk}=0$ otherwise. In our definition we have no $\mathrm{i}$, because we chose $\weyl{1,1}=\mathrm{i}Y$. Therefore, the $\weyl{\xi}$ fulfill the commutation relations 
      \begin{equation}
        \label{eq:weyl_commute}
        \weyl{\xi}\weyl{\eta}=(-1)^{\xi_X\eta_Z-\xi_Z\eta_X}\weyl{\eta}\weyl{\xi}.
      \end{equation}
      In both cases terms of the type $\xi_X\eta_Z$ are scalar products where the addition is carried out modulo~$2$.\footnote{The definition of a scalar product is possible, because there are always only finitely many non-zero entries.} We call the function 
      \begin{equation}
        \sigma(\xi,\eta)=\xi_X\eta_Z-\xi_Z\eta_X
      \end{equation}
      the symplectic form. Thus, the commutation relations are encoded in the symplectic form.
            
      To further simplify our classical description, we transform the binary strings to Laurent polynomials using 
      \begin{equation}
        \label{eq:fourier}
        \hat \xi(\varia)=\sum_{x\in\ZZ}\xi(x)\varia^x.
      \end{equation}
      We use these polynomials as abstract objects without ever evaluating them for some value of $\varia$.
      Our example observable then looks like
      \begin{displaymath}
        \quad\left(\begin{array}{cccccccc}
                      \cdots&0&1&\ul{1}&0&1&0&\cdots\\
                      \cdots&0&0&\ul{1}&1&0&0&\cdots
                    \end{array}\right)
         \mapsto\binom{u^{-1}+1+u^2}{1+u}.
      \end{displaymath}
      We change the function ${\bf w}$ in a way that it maps vectors of Laurent polynomials to tensor products of Pauli matrices. By a slight abuse of notation, we will also name it ${\bf w}$.
      
      Now we want to describe CQCAs in this picture. A CQCA $\CQCA$ maps every Pauli matrix to a tensor product of those. Because it is an automorphism and determined by the local transformation $\CQCA_0$, we can calculate the image of any observable from the image of the Pauli matrices it can be decomposed into. Of course, the images of neighboring Pauli matrices in general overlap and have to be multiplied. Because we required them to commute, this multiplication is uniquely defined. In the phase-space picture with binary strings this multiplication becomes an addition. The CQCA's action is thus a convolution of the observable's vector of binary strings $\tbinom{\xi^t_X}{\xi^t_Z}$ with the automaton's single cell images $\tbinom{\sca_{X\to X}}{\sca_{X\to Z}}$ and $\tbinom{\sca_{Z\to X}}{\sca_{Z\to Z}}$:
      \begin{eqnarray*}
        \binom{\xi^{t+1}_X(x)}{\xi^{t+1}_Z(x)}&=&
        \left(\left(\begin{array}{cc}
                \sca_{X\to X}&\sca_{Z\to X}\\
                \sca_{X\to Z}&\sca_{Z\to Z}
              \end{array}
        \right)
        \bigstar\binom{\xi^t_X}{\xi^t_Z}\right)(x)\\
        &=&\binom{(\sca_{X\to X}\star\xi^t_X)(x)+(\sca_{Z\to X}\star\xi^t_Z)(x)}{(\sca_{X\to Z}\star\xi^t_X)(x)+(\sca_{Z\to Z}\star\xi^t_Z)(x)}.
      \end{eqnarray*}
      The transformation (\ref{eq:fourier}) to the Laurent polynomials has the nice property of turning convolutions into multiplications:
      \begin{eqnarray*}
        \widehat{\xi\star\eta}&=&\widehat{\sum_y\xi(-y)\tau_y\eta}\\
        &=&\sum_x\sum_y\xi(-y)\eta(x+y)\varia^x\\
        &=&\sum_{k=x+y}\sum_{l=-y}\xi(l)\eta(k)\varia^{k+l}\\
        &=&\sum_l\xi(l)\varia^l\sum_k\eta(k)\varia^k\\
        &=&\hat\xi\cdot\hat\eta.        
      \end{eqnarray*}
      Because of this property, we refer to (\ref{eq:fourier}) as an algebraic Fourier transform. We get     
      \begin{eqnarray*}
        \binom{\hat\xi^{t+1}_X(\varia)}{\hat\xi^{t+1}_Z(\varia)}&=&
        \left(\begin{array}{cc}
                \hat\sca_{X\to X}(\varia)&\hat\sca_{Z\to X}(\varia)\\
                \hat\sca_{X\to Z}(\varia)&\hat\sca_{Z\to Z}(\varia)
              \end{array}
        \right)
        \cdot\binom{\hat\xi^t_X(\varia)}{\hat\xi^t_Z(\varia)}\\
        &=&\binom{\hat\sca_{X\to X}(\varia)\cdot\hat\xi^t_X(\varia)+\hat\sca_{Z\to X}(\varia)\cdot\hat\xi^t_Z(\varia)}{\hat\sca_{X\to Z}(\varia)\cdot\hat\xi^t_X(\varia)+\hat\sca_{Z\to Z}(\varia)\cdot\hat\xi^t_Z(\varia)}.
      \end{eqnarray*}
      In the following, we will omit the hat ``$\hat{\phantom{a}}$'' and the variable $\varia$ for the sake of a short notation. Furthermore, we will replace $\xi_X$ by $\xi_+$, $\xi_Z$ by $\xi_-$, and $\sca_{X\to X}$ by $\sca_{11}$ etc.\ to be consistent with the notation introduced in \cite{SchlingemannCQCA}. We then have
      \begin{equation}
        \xi^{t+1}=\sca\xi^t=\scamat\binom{\xi_+^t}{\xi_-^t}.
      \end{equation}
      Our example glider CQCA now looks like this:
      \begin{equation}
        \label{eq:glider_cqca}
        \sca_G=\left(\begin{array}{cc}
          0&1\\
          1&\varia^{-1}+\varia
        \end{array}
        \right).
      \end{equation}
      
      We have already seen that the CQCAs have to fulfill certain conditions, namely the local rule has to be a translation-""invariant homomorphism which maps Pauli products to Pauli products. Furthermore, it has to obey commutation relations with its translates. We have to translate these conditions to the polynomial picture. The matrix we use does not have any dependence on the position on the chain, so translation invariance is already included in this formulation. The commutation relations are encoded in the symplectic form, as we can see from (\ref{eq:weyl_commute}). For $\CQCA$ to conserve the commutation relations, the corresponding classical automaton has to conserve the symplectic form. This is why the classical CAs that correspond to CQCAs are called symplectic cellular automata (SCAs). In \cite{SchlingemannCQCA} it was proven that a $2\times 2$ matrix $\sca$ with Laurent-polynomial entries is a SCA if and only if it fulfills the following conditions:
      \begin{enumerate}
        \item $\det(\sca)=\varia^{2a},\,a\in\ZZ$; 
        \item all entries $\sca_{ij}$ are symmetric polynomials centered around the same (but arbitrary) lattice point $a$;
        \item the entries $\sca_{1j}$, $\sca_{2j}$ of both column vectors, which are the pictures of $(1,0)$ and $(0,1)$, are coprime.
      \end{enumerate}      
      Furthermore, it was shown that to every CQCA there exists a SCA and an appropriate translation-""invariant phase function $\lambda(\xi)$ such that
      \begin{equation}
        \CQCA[\weyl{\xi}]=\lambda(\xi)\weyl{\sca\xi},
      \end{equation}
      \begin{equation*}
        \lambda(\xi+\eta)=\lambda(\xi)\lambda(\eta)(-1)^{\xi_+\eta_--(\sca\xi)_+(\sca\eta)_-}
      \end{equation*}
      and $|\lambda(\xi)|=1\;\forall\xi$ hold. Additionally, $\lambda(\xi)$ is uniquely determined for all $\xi$ by the choice of $\lambda$ on one site. On the other hand, we can find CQCAs for any given SCA by adding a phase function. Thus, CQCAs and SCAs are equivalent up to a phase. We will therefore only refer to CQCAs, even if we talk about the corresponding SCAs.
      
      The last condition, the homomorphism property, is automatically fulfilled because the choice of the phase function and the conservation of the symplectic form $\sigma(\sca\xi,\sca\eta)=\sigma(\xi,\eta)$. The multiplication of Pauli matrices is mapped to the addition (modulus two) of phase-space vectors. As our matrices are linear transformations they obey $\sca(\xi+\eta)=\sca\xi+\sca\eta$. This translates to $\CQCA([\weyl{\xi}\weyl{\eta}]=\CQCA[\weyl{\xi}]\CQCA[\weyl{\eta}]$ via
      \begin{eqnarray*}
        &&\CQCA[\weyl{\xi}\weyl{\eta}]\\
        &=&(-1)^{-\eta_+\xi_-}\CQCA[\weyl{\xi+\eta}]\\
        &=&\lambda(\xi+\eta)\weyl{\sca(\xi+\eta)}(-1)^{-\eta_+\xi_-}\\
        &=&(-1)^{(\sca\eta)_+(\sca\xi)_-}\weyl{\sca\xi}\weyl{\sca\eta}\lambda(\xi+\eta)(-1)^{-\eta_+\xi_-}\\
        &=&\lambda(\xi)^{-1}\lambda(\eta)^{-1}\CQCA[\weyl{\xi}]\CQCA[\weyl{\eta}]\lambda(\xi)\lambda(\eta)\\ &&\cdot(-1)^{\xi_+\eta_--(\sca\xi)_+(\sca\eta)_-}(-1)^{-\eta_+\xi_-+(\sca\eta)_+(\sca\xi)_-}\\
        &=&\CQCA[\weyl{\xi}]\CQCA[\weyl{\eta}](-1)^{\sigma(\xi,\eta)-\sigma(\sca\xi,\sca\eta)}\\
        &=&\CQCA[\weyl{\xi}]\CQCA[\weyl{\eta}].
      \end{eqnarray*}
      
      A very simple CQCA is the shift on the lattice, which has the matrix $\varia^a\Id$. It obviously commutes with all other CQCAs. Thus, we can multiply a CQCA with determinant $u^{2a}$ which has entries that are centered around $a$ by a shift by $-a$ sites to obtain a centered CQCA with determinant $1$. From now on we will only consider centered CQCAs. 
      
      A nice property of these centered CQCAs, which we will need to prove that the entanglement generation is linear, is that their matrices form a multiplicative group. Multiplying the matrices means concatenating the CQCAs. 
    \subsubsection{Classes of CQCAs}
\begin{figure}[tp]
        \begin{center}
          \scalebox{1}{
            \includegraphics[angle=180,width=\columnwidth]{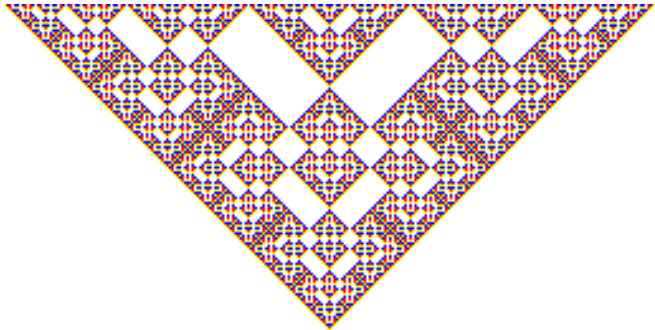}
          }
        \caption{Time evolution of the fractal CQCA $\CQCA_F$ with $\tr\sca_F=\varia^{-1}+1+\varia$. Time increases upwards. The different colors mark the different Pauli matrices.}
        \label{fig:fractal}
        \end{center}
      \end{figure}

      CQCAs show a variety of time evolutions that can be roughly grouped into three classes. The first class shows periodic behavior, the second class consist of CQCAs that have glider observables, which just move on the lattice as shown in Figure~\ref{fig:glider}, and the last case generates fractal space-time evolutions as shown in Figure~\ref{fig:fractal}. The class of a CQCA is determined by the trace of its matrix. If the trace is a constant $\tr\sca=c,\,c\in\ZZ_2$, the automaton is periodic. If it is of the form $\tr\sca=\varia^{-n}+\varia^n,\,n\in\NN$, the CQCA has gliders that move $n$ steps on the lattice every time step. All other cases show fractal behavior \cite{Guetschow2009}. We have already seen an example of a glider CQCA. The following CQCA exhibits fractal behavior:
      \begin{equation}
        \sca_F=\left(\begin{array}{cc}
          \varia^{-1}+1+\varia&\;1\\
          1&0
        \end{array}
        \right).
      \end{equation}
      Its trace is $\tr\sca_F=\varia^{-1}+1+\varia$. The time evolution that is shown in Figure~\ref{fig:fractal} is very similar for all CQCAs with the same trace. 
      This is observed also for other classes and subclasses of CQCAs. Mathematically, however, in general there is no notion of equivalence known for CQCAs with the same trace. An exception are CQCAs with trace $\varia^{-1}+\varia$, which can be shown to be all equivalent in the sense that they can all be transformed into the standard glider CQCA (\ref{eq:glider_cqca}) via
      \begin{equation}
        \sca=\scb\sca_G\scb^{-1}
      \end{equation}
      where $\scb$ is a CQCA \cite{Guetschow2009}. 
\section{Pure Translation-Invariant Stabilizer States}
  \subsection{Definition}
    \label{sec:stab_def}
    A stabilizer state $\omega$ is a common eigenstate to a group of commuting operators $\stabilizer=\left<\{S_i\}\right>$. This means that $\omega\circ S=\omega,\quad\forall S\in \stabilizer$. The group is generated by the set of generators $\stabgen=\{S_i\}$ by multiplication. It therefore suffices to check the stabilizer condition $\omega\circ S=\omega$ for the generators $S_i$. For finitely many qubits, this can be easily understood with the following example:
    \begin{example}
      \label{exam:bell}
      The stabilizer group with the generators $\stabgen=\{X\otimes X,Z\otimes Z\}$ stabilizes the Bell state $\psi=1/\sqrt{2}(\ket{1,1}+\ket{0,0})$. To check this, we just have to apply the stabilizer generators to the state (we omit the normalization):
      \begin{eqnarray*}
        (X\otimes X)\psi&=&(X\otimes X)(\ket{1}\otimes\ket{1})+(X\otimes X)(\ket{0}\otimes\ket{0})\\
        &=&\ket{0,0}+\ket{1,1}=\psi\\
        (Z\otimes Z)\psi&=&(Z\otimes Z)(\ket{1}\otimes\ket{1})+(Z\otimes Z)(\ket{0}\otimes\ket{0})\\
        &=&(-\ket{1})\otimes(-\ket{1})+\ket{0}\otimes\ket{0}\\
        &=&(-1)^2\ket{1,1}+\ket{0,0}=\psi
      \end{eqnarray*}
      \fin
    \end{example}

    For infinitely many qubits we can't just write down the state, so we use an abstract definition. A translation-""invariant stabilizer state is defined by a translation-""invariant set of operators $\stabgen=\{\weyl{\tau_x\xi},\,x\in\ZZ\}$, where $\tau_x$ is the lattice translation by $x$ sites. In \cite{SchlingemannCQCA} it was proven that such a set defines a pure translation-""invariant stabilizer state if and only if $\xi$ is reflection invariant and the Laurent polynomials of $\xi$ have no common divisors: $\gcd(\xi_+,\xi_-)=1$. For the polynomials, this means that
    \begin{enumerate}
      \item $\xi$ is of odd length, because reflection-invariant Laurent polynomials of even length are always divisible by $1+\varia$. We will write $l=2n+1$. 
      \item $\xi(0)\ne \binom{0}{0}$; thus, the center element is not $\Id$. Else $\xi$ would have the divisor $\varia^{-1}+\varia$.
      \item At least two different types of Pauli matrices (both different from the identity) have to occur (e.g. $X$ and $Y$). Otherwise, $\xi_+=0$ or $\xi_-=0$ or $\xi_+=\xi_-$, each case implying common divisors.
    \end{enumerate}
    An example is $S_i=X_{i-1}\otimes Z_i\otimes X_{i+1}$. The condition that the polynomials of $\xi$ are coprime is also a condition for the column vectors of a CQCA matrix. This indicates a close connection between CQCAs and transla-tion-""invariant stabilizer states. Indeed, CQCAs map pure translation-""invariant stabilizer states onto each other. Furthermore, every translation-""invariant stabilizer state can be generated by a single step of a CQCA from the ``all-spins-up'' state, which is a stabilizer state with stabilizer generators $\stabgen=\{\weyl{\tau_x(0,1)}=Z_x,\,x\in\ZZ\}$. Thus, we can study the entanglement generation properties of CQCAs acting on pure translation-""invariant stabilizer states by just studying the entanglement in these states. 
  \subsection{Entanglement}
    Entanglement in the stabilizer formalism was first studied in \cite{Fattal2004}. It was found that the entropy of entanglement with respect to a bipartite cut of some stabilizer state is exactly the number of generators of the correlation subgroup. The correlation subgroup $\stabilizer_{AB}$ for a bipartite stabilizer state is generated by all minimal stabilizer generators that have support on both parts of the system. The minimal stabilizer generators are those stabilizer operators which generate the stabilizer group (through multiplication) while having the smallest support of all such sets of operators. There are in general a lot of such sets, so the choice is not unique. For example, the stabilizer generators $Z\otimes\Id$ and $\Id\otimes Z$ would stabilize the product state $\psi=\ket{\uparrow}\otimes\ket{\uparrow}$. Obviously the combination $Z\otimes Z$ and $\Id\otimes Z$ would do the same, but with one generator with larger support. In this case the first set would be minimal, while the second wouldn't. In the case of translation-""invariant pure stabilizer states we don't have this issue. The set of generators is always translation invariant and fulfills the conditions introduced in Section~\ref{sec:stab_def}. Assume that a given generator of a translation-invariant pure stabilizer state is not minimal. Then it is composed of at least two stabilizer generators which also have to fulfill the conditions from Section~\ref{sec:stab_def}. In particular, all of the generators have to be the same. This implies that the polynomials of the original non-minimal generator have common divisors. But this is not possible, because it is required that the polynomials are coprime. Thus the generators of translation-invariant pure stabilizer states are always minimal.
    
    Unfortunately the proof in \cite{Fattal2004} relies heavily on the fact that only finite systems are considered. We use a different approach and come to essentially the same result that holds for translation-""invariant stabilizer states on infinite chains.
    \subsubsection{The Bipartite Case}
\begin{figure}[htp]
        \begin{center}
          \scalebox{1}{
            \includegraphics{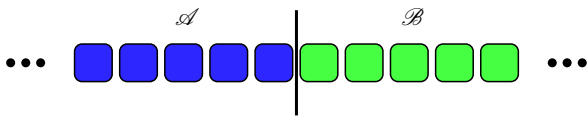}
          }
        \caption{The spin chain is cut into two halfchains $\SCA$ and $\SCB$. We study the entanglement between these halfchains.}
        \label{fig:halfchains}
        \end{center}
      \end{figure}

      First we will investigate the case of a bipartite splitting of the chain. We have two parties, say Alice and Bob, where Alice controls the part $\SCA$ and Bob the part $\SCB$ of the chain. This is shown in Figure~\ref{fig:halfchains}. 
      Let us first define bipartite entanglement for stabilizer states.
      \begin{definition}
        The entanglement $E(\omega_\xi)$ of a translation-""invariant stabilizer state $\omega_\xi$ in a bipartite setting is the number of maximally entangled qubit pairs with respect to any bipartite cut. These qubit pairs are logical ones, that is they are each localized on several physical qubits. By local operations on each part of the chain one could localize the logical qubits onto one physical qubit each.
      \end{definition}
      
      In this case we have the following theorem:
      \begin{thm}
        \label{thm:stab_ent}
          A pure translation-""invariant stabilizer state of stabilizer generator length $2n+1$ entangles $n$ qubit pairs maximally with respect to any bipartite cut.
        \begin{proof}
          Here we will only present the idea of the proof. The technical parts can be found in the appendix of \cite{Guetschow2009}
        
          Obviously, all bipartite cuts are equivalent as the state is translation-""invariant. Thus, we can look at any particular cut to prove the general result. We have a stabilizer generator centered around each site. Unless our stabilizer generators are single site operators ($n=0$), the cut will always leave several stabilizer generators cut into parts on both systems $\SCA$ and $\SCB$. As one can see in Figure~\ref{fig:cut_stab} there will be $2n$ stabilizer generators affected. These operators generate the correlation subgroup $\stabilizer_{AB}$. Thus, the correlation subgroup has $2n$ generators and therefore dimension $|\stabilizer_{AB}|=2n$. The interesting fact about these stabilizer generators is that, despite commuting as a whole, their restrictions to $\SCA$ or $\SCB$ don't necessarily commute. Now we try to find commuting pairs of anticommuting Pauli products in the restriction of $\stabilizer_{AB}$ to $\SCA$ or $\SCB$ by multiplying stabilizer generators from the correlation subgroup (We check for the anti-commuting parts only on one side, but carry out the multiplication on both sides to preserve the stabilized state). We know from the theory of quantum error correction codes that each such pair on $\SCA$ encodes one qubit \cite{Gottesman1997}. Because we carried out the multiplication on both sides, the corresponding parts of the operators on $\SCB$ fulfill the same commutation relations and therefore also encode qubits. The pairs of operators on each halfchain behave like pairs of $X$ and $Z$; thus, we call them $\bar X$ and $\bar Z$.  We only required them to anti-commute, so we did not fix which of them is the $X$ and which the $Z$. Thus, we can choose this and we choose it in such a way that the corresponding operators on each side are either bot $\bar Z$ or both $\bar X$. Thus, we have a pair of stabilizer generators that reads $\bar X_\SCA\otimes\bar X_\SCB$ and $\bar Z_\SCA\otimes\bar Z_\SCB$. But, as seen in Example \ref{exam:bell}, $\stabgen=\{X\otimes X,Z\otimes Z\}$ encodes a Bell state, which is maximally entangled. Thus, each pair of anticommuting pairs stabilizes a maximally entangled (logical) qubit pair.
          \begin{figure}[htp]
            \begin{center}
              \scalebox{1}{
                \includegraphics{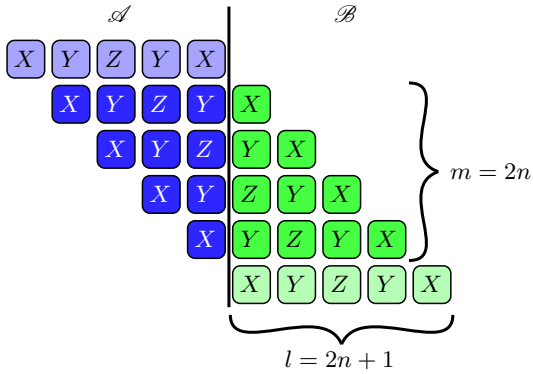}
              }
              \caption{Cut stabilizer generators: it is apparent that for operators of length $l=2n+1$, $2n$ of them have support on both $\SCA$ and $\SCB$.}
              \label{fig:cut_stab}
            \end{center}
          \end{figure}
        
          What is left to show is that we can always find $n$ such pairs. The proof is rather lengthy and technical. It is carried out in \cite{Guetschow2009} and based on methods from quantum error correction codes to directly construct the pairs. 
        \end{proof}
      \end{thm}
    \subsection{The Tripartite Case}
\begin{figure}[htp]
        \begin{center}
          \scalebox{1}{
            \includegraphics{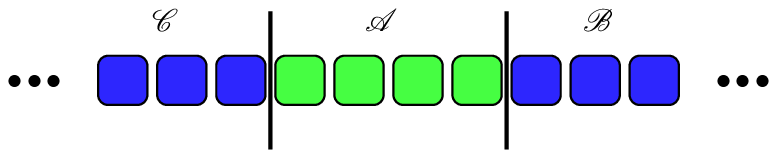}
          }
        \caption{A finite region $\SCA$ of $4$ spins is cut out of the chain leaving two infinite ends $\SCB$ and $\SCC$.}
        \label{fig:finite_region}
        \end{center}
      \end{figure}

      In this setting we cut the chain into three parts, one middle part of length $L$ and two infinite ends, as shown in Figure~\ref{fig:finite_region}.
      We now want to calculate the entanglement between the finite part and the two infinite parts. To do this calculation, we use the same method as above, and arrive at the following theorem.
      \begin{thm}
        \label{thm:stab_ent_encode_finite}
        Given a pure translation-""invariant stabilizer state $\omega_\xi$ of stabilizer generator length $2n+1$, a region of length $L$ shares $2n$ maximally entangled qubit pairs with the rest of the chain if $2n\le L$ and $L$ qubits pairs if $2n>L$.
        \begin{proof}
          The proof works exactly as in the bipartite case. In the case $2n\le L$ the cut stabilizers are only cut on one side. But all stabilizers that are cut on the left-hand side commute with those cut on the right-hand side. Thus, we have two independent cuts of the bipartite case and therefore $2n$ pairs of maximally entangled qubits. In the case $2n > L$ some stabilizers are cut on both sides. We use the same technique to produce the mutually commuting anti-commuting pairs which encode the qubits as in Theorem \ref{thm:stab_ent}. We always find $L$ pairs of maximally entangled qubits.
        \end{proof}
      \end{thm}
\section{Entanglement Generation}
\begin{figure}[htp]
    \begin{center}
      \scalebox{1}{
        \includegraphics[width=\columnwidth]{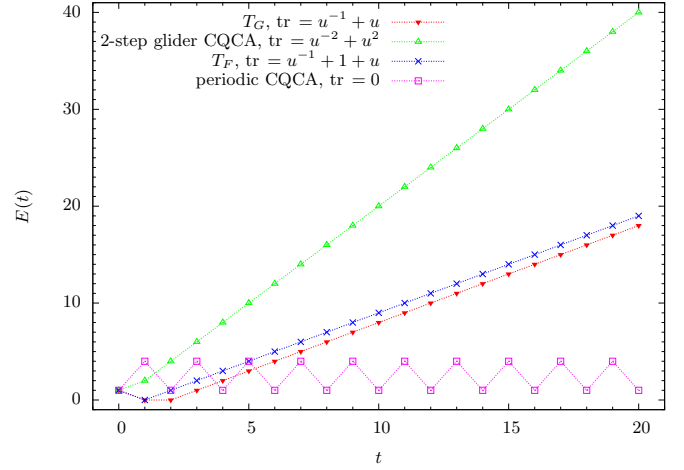}
      }
      \caption{Entanglement generation for the stabilizer state with $\weyl{\xi}=Y\otimes X\otimes Y$ in a bipartite setting with different CQCAs. One can see that entanglement can also be destroyed, but grows asymptotically linearly with the number of time steps $t$. The rate is given by the degree of the trace of the CQCA matrix.}
      \label{fig:bipartite_stab}
    \end{center}
  \end{figure}

  Now we come to the generation of entanglement through CQCA action. As we have seen in the previous section, the bipartite entanglement of a pure translation-""invariant stabilizer state depends linearly on the length of the generators of the stabilizer group. So, it suggests itself to study the evolution of the length of the stabilizer generators under CQCA action. First, let us define the asymptotic entanglement generation rate from stabilizer states.  
  \begin{definition}
    The asymptotic entanglement generation rate from stabilizer states for CQCAs is defined as 
    \begin{equation}
      \frac{\Delta E}{\Delta t}(\omega_\xi) =\lim_{t\to\infty}\frac{1}{t}E(\omega_\xi,t),
    \end{equation}
    where $E(\omega_\xi,t)$ is the bipartite entanglement at time~$t$.
  \end{definition}
  \begin{lemma}
    \label{lem:stab_growth}
    The length $2n+1$ of the minimal stabilizer generators grows asymptotically with
      \begin{equation}
        \frac{\Delta n}{\Delta t}(\omega_\xi) =\lim_{t\to\infty}\frac{1}{t}(2n(\omega_\xi,t)+1)=\dg{\tr\sca}
      \end{equation}
    for any centered CQCA $\CQCA$ and any translation-invariant pure stabilizer state $\omega_\xi$.
    \begin{proof}
      We know that CQCAs map pure translation-""invariant stabilizer states onto pure translation-""invariant stabilizer states. The image of a state with stabilizer generators $\stabgen=\{\weyl{\tau_x\xi},\,x\in\ZZ\}$ under the action of $t$ steps of a CQCA $\CQCA$ is a state with stabilizer generators $\stabgen^t=\{\weyl{\tau_x\sca^t\xi},\,x\in\ZZ\}$. Furthermore, we know that any stabilizer state can be generated from the ``all-spins-up'' state by a CQCA $\scb$. So we have $\stabgen^t=\{\weyl{\tau_x\sca^t\scb(0,1)},\,x\in\ZZ\}$. The length of the stabilizer generator is determined by the highest order of the stabilizer generator polynomials, $\dg{\xi}$. Namely, the stabilizer generator is of length $2\cdot\dg{\xi}+1$. So, we have to calculate $\dg{\sca^t\xi}=\dg{\sca^t\scb\tbinom{0}{1}}$.
    
      For an arbitrary product of CQCAs $\prod_{i=1}^k\sca_i$ we can define the series $(a_l)_{1\le l\le k}=\dg{\prod_{i=1}^l\sca_i}$. It is subadditive, i.e. $a_{n+m}\le a_n+a_m$, because the concatenation of CQCAs is essentially the multiplication and addition of polynomials, which is subadditive in the exponents. For subadditive series $a_n$ Fekete's lemma \cite{Fekete1923} states that the limit $\lim_{n\to\infty}\frac{a_n}{n}$ exists. In our case the series is always positive, so the limit is positive and finite. An easy way to determine the limit is to take a subseries, which of course has the same limit. The subseries of the $t=2^k$th $(k\in\NN)$ steps is a good candidate, because we can make use of the Cayley-Hamilton theorem to obtain
      \begin{equation*}
        \sca^{2^k}=\sca(\tr\sca)^{2^k-1}+\Id\sum_{i=1}^k(\tr\sca)^{2^k-2^i}.
      \end{equation*}
      Furthermore, we have
      \begin{eqnarray*}
        \dg{\sca^{2^k}\scb\tbinom{0}{1}}&=&\dg{\sca\scb\tbinom{0}{1}(\tr\sca)^{2^k-1}+\scb\tbinom{0}{1}\sum_{i=1}^k(\tr\sca)^{2^k-2^i}}\\
        &=&a\cdot c(k)+b\cdot d(k),
      \end{eqnarray*}
      with $a=\sca\scb\tbinom{0}{1}$, $b=\scb\tbinom{0}{1}$, $c(k)=(\tr\sca)^{2^k-1}$, $d(k)=\sum_{i=1}^k(\tr\sca)^{2^k-2^i}$.
             
      Let us first assume that for some $k_0$ we have $\dg{\sca^{2^k}\scb\tbinom{0}{1}}>\dg{b}$. We start by determining a recursion relation for $c(k)$ and $d(k)$.
      \begin{eqnarray*}
        c(k+1)&=&(\tr\sca)^{2^{k+1}-1}=(\tr\sca)^{2^k-1}(\tr\sca)^{2^k}\\
        &=&c(k)(\tr\sca)^{2^k},\\
        d(k+1)&=&\sum_{i=1}^{k+1}(\tr\sca)^{2^{k+1}-2^i}\\
        &=&\sum_{i=1}^{k}(\tr\sca)^{2^k+2^k-2^i}+(\tr\sca)^0\\
        &=&d(k)(\tr\sca)^{2^k}+1.
      \end{eqnarray*}
      Now we are able to calculate the limit:
      \begin{eqnarray*}
        &&\lim_{k\to\infty}\frac{1}{2^k}\dg{\sca^{2^k}{\bf b}\tbinom{0}{1}}\\
        &=&\lim_{k\to\infty}\frac{1}{2^k}\dg{a\cdot c(k)+b\cdot  d(k)}\\
        &=&\lim_{k\to\infty}\frac{1}{2^k}\dg{a\cdot c(k-1)(\tr\sca)^{2^{k-1}}\\
        &&+b\cdot d(k-1)(\tr\sca)^{2^{k-1}}+b}\\
        &=&\lim_{k\to\infty}\frac{1}{2^k}\dg{(\tr\sca)^{2^{k-1}}r(k-1)}\\
        &=&\lim_{k\to\infty}\frac{1}{2^k}\left(\dg{\tr\sca}\frac{2^{k}}{2}+\dg{r(k-1)}\right)\\
        &=&\lim_{k\to\infty}\left(\sum_{i=1}^{k-k_0}\frac{1}{2^i}\dg{\tr\sca}\right)+\underbrace{\lim_{k\to\infty}\frac{1}{2^k}\dg{r(k_0)}}_{=0}\\
        &=&\dg{\tr\sca}.
      \end{eqnarray*}
      In the third step we used that $\dg{b}<\dg{\sca^{2^k}\scb\tbinom{0}{1}},\,\forall k\ge k_0$.
        
      In the second case we have $\dg{\sca^{2^k}\scb\tbinom{0}{1}}\le\dg{b},\,\forall k$. Thus, $\dg{\sca^{2^k}\scb\tbinom{0}{1}}$ is bounded by $\dg{b}$. Therefore, $n(t,\xi)$ is bounded and $\frac{\Delta n}{\Delta t}=0$. But $n(t,\xi)$ bounded also implies $\sca$ periodic and therefore $\dg{\tr\sca}=0$. Thus, we have $\frac{\Delta n}{\Delta t}=0=\dg{\tr\sca}$. This completes the proof.
    \end{proof}
  \end{lemma}
  Now we only have to put Theorem~\ref{thm:stab_ent} together with Lemma~\ref{lem:stab_growth} to prove the following theorem:  
  \begin{thm}
    The asymptotic bipartite entanglement generation rate (maximally entangled qubit pairs per step) of a general centered CQCA $\CQCA$ is the degree of its trace polynomial, $\dg{\tr\sca}$.
    \begin{proof}
      By Theorem~\ref{thm:stab_ent}, every translation-""invariant stabilizer state with stabilizer generator length $2n+1$ entangles $n$ qubit pairs maximally with respect to any bipartite cut. Lemma \ref{lem:stab_growth} shows that under the action of a CQCA $\CQCA$ the length of the stabilizer generators grows asymptotically with $\dg{\tr\sca}$ qubit pairs per step. This implies the proposition.
    \end{proof}
  \end{thm}
  This means that starting from the ``all-spins-up'' stabilizer product state, the entanglement grows linearly with $\dg{\tr\sca}$ under the action of a CQCA $\CQCA$, because $\scb=\Id$. Starting from an arbitrary translation-""invariant pure stabilizer state, the entanglement might decrease in the first $k$ steps, e.g.\ if $\sca^k=\scb^{-1}$, but then starts to increase linearly with $\dg{\tr\sca}$. This behavior is shown in Figure~\ref{fig:bipartite_stab}. 
  
\begin{figure}[htb]
    \begin{center}
      \includegraphics[width=\columnwidth]{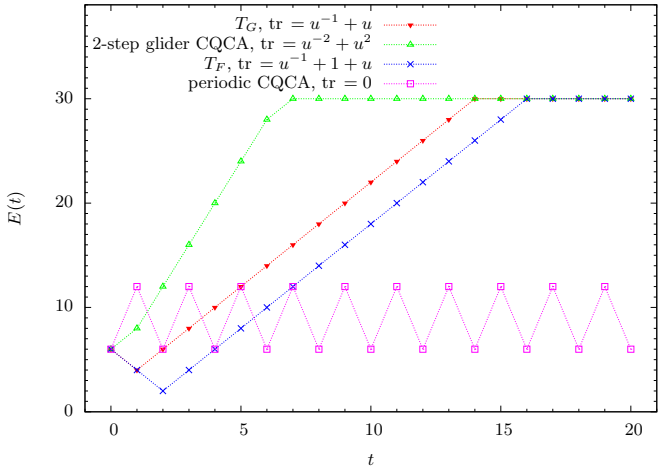}
      \caption{Evolution of entanglement of a subchain of $30$ consecutive spins for an initial  stabilizer state with $\weyl{\xi}=Y\otimes X\otimes X\otimes X\otimes X\otimes X\otimes Y$ for different CQCAs. The entanglement first grows as in the bipartite case, but then saturates at $30$ qubit pairs.}
      \label{fig:tripartite_stab}
    \end{center}
  \end{figure}

  In the tripartite setting, again starting from the ``all-spins-up'' state, the entanglement between the finite region and the rest of the chain grows under the action of a CQCA $\CQCA$ with $2\cdot\dg{\tr\sca}$ until it reaches $L$. Then it remains constant. If we start with a general translation-""invariant stabilizer state, again the entanglement might decrease at first. After some time it starts increasing and reaches $L$, where it remains if the CQCA is not periodic. The entanglement generation is twice as fast as for the bipartite setting, because we have two cuts. Results are shown in Figure~\ref{fig:tripartite_stab}.
  
  In both cases, the asymptotic entanglement generation only depends on the degree of the trace of the automaton. A CQCA with gliders generates entanglement just as fast as a fractal automaton if the degree of the trace is the same. Of course, periodic automata destroy all entanglement they just generated in the next steps. This corresponds to the fact that their trace is a constant. The entanglement generation rate is however not directly governed by the neighborhood of the CQCA. Of course, the size of the neighborhood bounds the possible rate of entanglement generation from above, but even automata with a huge neighborhood can be periodic and thus generate no entanglement at all.
  
  It is worth mentioning that CQCAs saturate the bound on the entanglement generation rate for translation-invariant operations acting on translation-invariant states derived in \cite{Guetschow2009}. This means that there is no translation-invariant operation that generates more entanglement per step, while having the same size of neighborhood, than a CQCA whose trace has maximal degree with respect to the neighborhood.
\section{Notes on Finite Chains}
  In our whole analysis we used an infinite spin chain to obtain translation invariance, which would be broken by the ends of the spin chain. For finite chains we generally have to take the effects of the ends into account. But when we deal with observables which are localized far away from the ends of the chain and time periods $T$ which are much smaller than the length of the chain, for locality reasons the time evolution of the observables has to be the same as for the infinite chain. Information from the ends of the chain can only travel with the same finite speed, as all other information, so e.g.\ a one-site observable localized in the middle of a chain of length $L=2l+1$ can only interfere with influences from the ends after $l/2$ time steps.
  
\begin{figure*}[tp]
    \begin{center}
      \scalebox{1}{
        \includegraphics{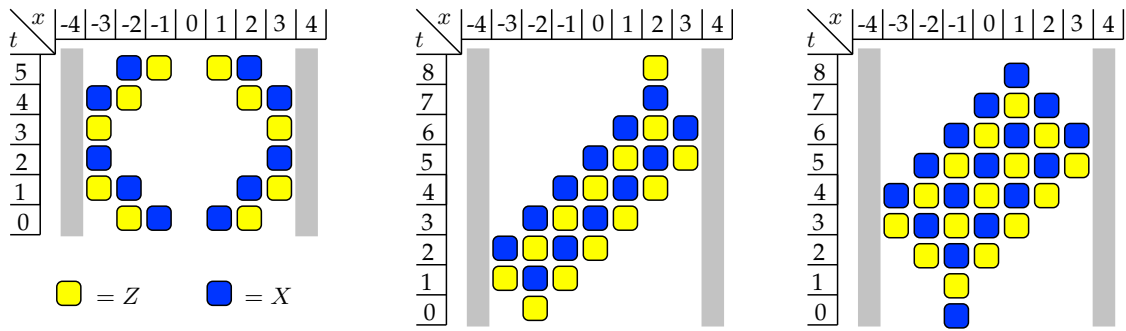}
      }
      \caption{Time evolution of the glider CQCA on a finite chain of $7$ spins. In the left-hand illustration we can see that the gliders are reflected at the ends of the chain. In the other two illustrations we see that on a finite chain the glider CQCA mirrors the position of single-site Pauli matrices. As all Pauli products can be decomposed into single-site Pauli matrices, this holds for all observables. (All observables can be decomposed into a sum of Pauli products, which are each mirrored in the same number of time steps, so the whole observable is mirrored.)}
      \label{fig:finite_glider}
    \end{center}
  \end{figure*}

  Admittedly the applications of CQCAs make heavy use of the effects occurring at the ends of the spin chains. Thus, our above statement is not applicable. To take effects at the ends of the spin chains into account, we have to deal with the fact that they break the translation invariance. The one-site images of the sites at the ends of the chain whose neighborhoods would reach over one end of the chain, if they were the same as on the rest of the chain, have to be adapted. Like in the case of stabilizer states above, in general the cut images don't fulfill the necessary commutation relations any more. There is no general theory yet of how to adapt the CQCAs at the ends of the spin chain. However, for special cases the cut images still fulfill the commutation relations. Fortunately, the much used glider CQCA is of this type. So, in this case the only influence from the ends of the chain is that the outermost sites miss the influence from one neighbor. As we still have an automorphism and thus a reversible operation on the whole chain, no information can be lost at the ends of the chain. So, the ends have to be reflective. We can observe that in the case of an incoming glider in the left part of Figure~\ref{fig:finite_glider}.   
  This leads to the fact that observables are mirrored by the chain. For single-site Pauli matrices, this can be seen in the right-hand part of Figure~\ref{fig:finite_glider}. Arbitrary observables are sums of tensor products of single-site Pauli matrices. Because the single-site Pauli matrices are mirrored, the tensor products will be mirrored, too. Thus, also sums are mirrored and therefore any observable is mirrored on the chain. This property and the spreading of observables in this process is used in applications of CQCAs like \cite{fitzsimons2006,Raussendorf2005a}.
\section{Applications}
  CQCAs are used in different quantum computational schemes. Of course they are not capable of doing any computations efficiently; a classical computer couldn't do efficiently as well, because they are essentially classical cellular automata. But, nevertheless they can be of great use when accompanied by non-Clifford operations. This is used in different approaches.
  \subsection{Measurement Based Quantum Computation}
    The most famous model of quantum computation involving a CQCA (indirectly) is the idea of measurement based quantum computation or the ``one way quantum computer'' by Raussendorf and Briegel \cite{Raussendorf2001}. The resource state used to perform the quantum computation by successively measuring the single qubits is a so-called cluster state on a 2-dimensional (finite) lattice of qubits. It can be generated by a CQCA, namely a 2-dimensional version of the glider CQCA (\ref{eq:glider_cqca}). Thus, one time-step of a CQCA together with measurements of single qubits suffices for universal quantum computation.
  \subsection{Raussendorf's Scheme of Translation Invariant Quantum Computation}
    In this scheme of translation-""invariant quantum computation \cite{Raussendorf2005a}, again the glider CQCA is used, in this case the one-dimensional version. The property of generating patterns from single-site observables, thus spreading them over the spin-chain (a finite chain is used here), is used to immunize observables against special global transformations. The time steps in which certain observables are immune against these operations depend on their initial position on the chain. Thus, temporal control (when to apply the global gates) can be turned into spatial control and any quantum operation can be conducted on the system. An example of such behavior would be the effect of a global Pauli $Y$ operation (a local $Y$ on each site). It gives a sign on each $X$ and $Z$ tensor factor. So, in our example shown in Figure~\ref{fig:finite_glider} the $Z_{-2}$ would gain a phase of $-1$ when the $Y$ is applied in one of the steps $\{0,1,6,7,8\}$, while it would gain no phase in the other steps. The $X_{-1}$ gains the phase in steps $\{0,1,2,3,6,7,8\}$. The steps where a phase is gained belong to the contraction resp.\ expansion of the observables, while the steps where no phase is gained belong to the transmission of the expanded observable over the chain. So, if we e.g.\ run the automaton for $2L+2=16$ time-steps and apply a global $Y$ in step $3$, $X_{-1}$ gains a phase but $Z_{-2}$ does not. Applying $Y$ in steps $\{1,3,6,7,\}$ $Z_{-2}$ gets the phase. So, we turned temporal control into spatial control. To achieve universal quantum computation, we use arbitrary translation-invariant local rotations instead of the global $Y$.
  \subsection{The Quantum Computational Scheme by Fitzsimons and Twamley}
    In \cite{fitzsimons2006} the same property of the glider CQCA is used. However in this case the non-Clifford operations are not translation-""invariant, but only conducted at the ends of the chain. Separate control of the ends of the chain is justified by the fact that due to the missing neighbor the physical properties of the systems at the end differ from those in the middle of the chain. The CQCA is used to transport qubits to the end of the chain, which can then be manipulated. Two-qubit gates are achieved by first decoupling one spin (at the end of the chain) from the chain and then transporting the other one to its neighboring position. Then the gate is applied and the qubits are transported back. This scheme was experimentally realized in a NMR-system \cite{fitzsimons2007}.
\section*{Conclusion and Outlook}
  We have introduced CQCAs and their classical description. It was shown that CQCAs acting on translation-invariant stabilizer states generate entanglement with the highest possible rate for translation-invariant operations. The rate only depends on the trace of the CQCA's matrix. It is independent of the class of the CQCA. We furthermore commented on some applications of CQCAs in quantum computational schemes. These schemes use finite chains of qubits, while our analysis is based on infinite spin-chains. Thus, a future task is, to complete the theory of CQCAs to include finite spin chains. This was already done for periodic boundary conditions \cite{SchlingemannCQCA}, but the case of non-periodic boundaries is still an open question. Additionally, the investigation of invariant states and convergence of states under CQCA action which was started in \cite{Guetschow2009} will be continued for finite chains to strengthen the connection to the applications of CQCAs.

\section*{Acknowledgements}
  The author would like to thank the Rosa Luxemburg Foundation for support. The original publication is available at www.springerlink.com (DOI: 10.1007/s00340-009-3840-1).

\bibliography{entanglement_generation_of_cqca}

\end{document}